\documentstyle[12pt]{article}

\begin{document}

\thispagestyle{empty}
\def\pubnum{344}
\def\data{September, 1994}

\begin{flushright}
{\parbox{3.5cm}{
UAB-FT-344

September, 1994

hep-ph/9410311
}}
\end{flushright}

\vspace{1cm}
\hyphenation{ne-ver-the-less}
\hyphenation{cano-nical ca-nonical canoni-cal}
\begin{center}
\begin{large}
\begin{bf}
SUPERSYMMETRIC ELECTROWEAK RENORMALIZATION OF THE $Z$-WIDTH IN THE MSSM (II)\\
\end{bf}
\end{large}
\vspace{0.25 cm}
David GARCIA
\footnote{Internet address:GARCIA@IFAE.ES}\,,
Ricardo A. JIM\'ENEZ
\footnote{Internet address:ABRAHAM@IFAE.ES}\,,
Joan SOL\`A
\footnote{Internet addresses:IFTESOL@CC.UAB.ES and SOLA@IFAE.ES}

\vspace{0.25cm}
Grup de F\'{\i}sica Te\`orica\\

and\\

Institut de F\'\i sica d'Altes Energies\\

\vspace{0.25cm}
Universitat Aut\`onoma de Barcelona\\
08193 Bellaterra (Barcelona), Catalonia, Spain\\
\end{center}
\vspace{0.2cm}
\begin{center}
{\bf ABSTRACT}
\end{center}
\begin{quotation}
\noindent
\hyphenation{ana-ly-ses a-na-ly-ses}
\hyphenation{de-li-ca-te}
\hyphenation{phe-no-me-no-lo-gi-cal}
\hyphenation{ge-nui-ne}

We address the computation of $\Gamma_Z$ and
of the intriguing quantity $R_b$ in the MSSM including full treatment of the
Higgs sector.
Contrary to previous partial approaches, and due to the possible relevance of
the
result to the fate of the MSSM,
we perform a complete calculation, without approximations.
For a pseudoscalar Higgs mass $m_{A^0}>70\,GeV$
and CDF limits on $m_t$,
the bounds on $R_b$ at $1\sigma$ level leave no room
to the MSSM to solve the ``$R_b$ crisis'' for any combination of the
parameters,
not even admitting the possibility of a light chargino and a light stop of
${\cal O}(50)\,GeV$; however, for $m_t$ not restricted by CDF,
a ``tangential''solution exists in the window
$2<\tan\beta<10$ with a light chargino and stop.
In contrast, for a pseudoscalar mass
$40\,GeV\stackrel{\scriptstyle <}{{ }_{\sim}} m_{A^0}<60\,GeV$ and CDF limits
on $m_t$,
the ``$R_b$ crisis''
can be solved in a comfortable way, for any SUSY spectrum above the
phenomenological
bounds, provided
$\tan\beta\stackrel{\scriptstyle >}{{ }_{\sim}} m_t/m_b$.
 Our general conclusion is that, if there is a ``$R_b$ crisis'' at all,
its solution within the MSSM has to do more with the peculiar structure of the
SUSY Higgs sector
rather than with the spectrum of genuine supersymmetric particles.
In view of the range predicted for $m_{A^0}$, LEP 200 should be able to
definitely
settle down this question.

\end{quotation}

\baselineskip=6.5mm

\newpage


Discovering Supersymmetry (SUSY)\,\cite{b1001}
would be a fact of paramount importance
both theoretically and phenomenologically in
the world of elementary particle physics.
In the past ten to fifteen years,
a lot of effort has being directed to settle down the question of whether SUSY
is real or not\,\cite{b1002}, and although the question remains unanswered
the quest  still goes on and on, even with renewed interest, especially with
the advent
of LEP 100, its subsequent planned upgrading to LEP 200,
and also spurred by the prospect of new and more powerful machines
in the future: LHC, $e^{+}e^{-}(500\,GeV)$,...
Whether the finding of SUSY particles--if real at all--  will have to await
physical production in those big collider
experiments, or perhaps some hints of existence might creep earlier
through non-negligible quantum effects on
physical observables, is not clear
for the moment, so we better keep on exploring both possibilities.
Here we shall
exploit one example of the second possibility, which may help to shed light
on SUSY physics through $Z$-decay dynamics. Indeed, at present
the cleanest and most accessible laboratory to test possible manifestations
of SUSY is LEP.
In Part\, I\,\,\cite{b1003}, whose notation and definitions we shall adopt
hereafter,
we have studied systematically the potential size of the full virtual
contributions
to the width of the $Z$ boson, $\Gamma_Z$, from the
plethora of ``genuine''($R$-odd) supersymmetric particles
of the MSSM; namely, from sleptons,
squarks, charginos and neutralinos,
with the result that for not too heavy sparticle masses (i.e. not heavier than
the
electroweak scale), they could be of the order or even larger than the
pure SM electroweak corrections--though opposite in sign in most cases.
This warns us of the possibility that there could
be a remarkable cancellation between the two contributions, and even an
overcompensation
of the (electroweak)
SM corrections by genuine SUSY effects. It also suggests to try to
better appreciate these features in particular decay channels, such as e.g. in
the
partial width of $Z\rightarrow b\bar{b}$, where the genuine SUSY contributions
are maximal.
However, to definitely assess whether this could be the case or not,
we have to take into account also the
additional contributions from the Higgs sector of the MSSM\,\cite{b1004,b1204}.
As in Part\, I, we identify the SM
with a ``Reference Standard Model'' (RSM)  in
which the Higgs mass is set equal to the mass of the lightest $CP$-even Higgs
scalar,  $h^0$,  of the MSSM. Since in a certain limit
 ($m_{A^0}\rightarrow\infty$, see later on)
the couplings of $h^0$  to fermions and gauge
boson are identical to those of the SM Higgs, we may easily subtract out the
RSM contribution from the MSSM.
In this way the total {\sl additional}
correction from the MSSM with respect to the RSM is given by
eq.(6) of Part\, I, viz.
\begin{equation}
\delta\Gamma_Z^{MSSM}=\delta\Gamma_Z^{H}+\delta\Gamma_Z^{SUSY}\,,
\label{eq:Tshift}
\end{equation}
where $\delta\Gamma_Z^{SUSY}$  has been considered in detail in that reference,
whereas $\delta\Gamma_Z^{H}$ will be taken into account in the present study.
Moreover, in this
note we shall also consider the full MSSM contribution (\ref{eq:Tshift}) in a
context where mixing effects in the third squark family are included
in the evaluation of $\delta\Gamma_Z^{SUSY}$. These effects were not considered
in
Part\, I, since we treated (conservatively) all squarks generations alike.
In the present case,
however, we will distinguish between the first two generations and
the last generation, where mixing effects are most likely to arise.
This will prove useful to emphasize the conclusion,
obtained from previous studies\,\cite{b1005,b1GHVR,b1BF5},
that the virtual SUSY corrections could help
in reducing the disagreement between theory and experiment for the
conflicting ratio\,\cite{b1006}
\begin{equation}
R_b^{\rm exp}={\Gamma_b\over\Gamma_h}\equiv {\Gamma (Z\rightarrow b\bar{b})
\over \Gamma (Z\rightarrow {\rm hadrons})}
=0.2208\pm 0.0024\,.
\label{eq:Rb}
\end{equation}
The SM prediction, including the variation with the top quark mass
within the allowed range of $\delta M_W$ and $\Delta r$,
is\,\cite{b1007}
\begin{equation}
R_b^{SM}=0.2158\pm 0.0013\,.
\label{eq:RbSM1}
\end{equation}
If one further incorporates the recently claimed CDF result
($m_t=174\pm 10\, ^{+13}_{-12}\,GeV$\,\cite{b1008}) for the top quark mass
it reads\,\cite{b1007}
\begin{equation}
R_b^{SM}|_{CDF}=0.2160\pm 0.0006\,.
\label{eq:RbSM2}
\end{equation}
It is well-known\,\cite{b1BH1} that the SM result decreases with $m_t^2$, due
to an
overcompensation of the propagator correction by a large, negative,
vertex contribution to the $b\bar{b}$ mode
\begin{equation}
\nabla_{V_b}^{SM} =\delta\rho_{V_b}-8|Q_b|s^2\,{v_b\,a_b\over v_b^2+a_b^2}\,
\delta\kappa_{V_b}
= -\,{4\over 3}\,{1+v_b/a_b\over 1+v_b^2/ a_b^2}\,\Delta\rho\simeq
-\,1.5\Delta\rho\,,
\end{equation}
where the dominant part of $\Delta\rho$ in the SM  is
\begin{equation}
\Delta\rho^t= {3\,G_F\,m_t^2\over 8\,\pi^2\,\sqrt{2}}\,.
\end{equation}
In comparing theory and experiment we shall consider the two
SM results, eqs.(\ref{eq:RbSM1})-(\ref{eq:RbSM2}), separately
\footnote{The announced ``evidence'' on the
top quark mass\,\cite{b1008} is for the moment not absolutely compelling and
we should be open-minded to all possible eventualities.}.
 In either case the discrepancy with
the experimental data is statistically significant: the SM prediction is
$\sim 2\sigma$ below the experimental result
\footnote{Although we are aware of the controversy over the measurement of
$R_b^{\rm exp}$ in connection to
$b$-tagging and its anticorrelation to $c$-tagging\,\cite{b11MM}, the matter is
not
settled at all. Thus we shall take the point of view that there is a ``$R_b$
crisis''
in the SM and explore its consequences in the MSSM. }.
Furthermore, the rather large preferred
CDF value for the top quark mass just goes in the opposite direction
to reconcile theory with experiment.
Fortunately, there is some hope to improve things in the framework of the MSSM,
where
to start with the fits to $m_t$ lead to a lighter central value
$m_t=162\pm 9\,GeV$\,\cite{b1EFL} compatible with the CDF errors,
whereas in the SM the central value
increases by as much as about $20\, GeV$, i.e. closer to the central CDF mass.
We shall therefore take advantage in our analysis
of the different values of $m_t$ at our disposal  and in  particular of the
favourable one corresponding to the MSSM fit.

Notice that the ratio $R_b$ is insensitive to $\alpha_s$. Moreover, since it is
also essentially independent of the Higgs mass in the SM,
we can simply identify the above theoretical predictions on
$R_b^{SM}$ with the RSM result, $R_b^{RSM}$.
Denoting by $\delta R_b^{MSSM}$ the radiative correction
induced on $R_b^{RSM}$ by the quantum effects (\ref{eq:Tshift}) on the various
partial widths, we have
\begin{equation}
R_b^{MSSM}=R_b^{RSM}+ \delta R_b^{MSSM}\,,
\label{eq:RbMSSM}
\end{equation}
where in an obvious notation
\begin{equation}
\delta R_b^{MSSM}=\delta R_b^{SUSY}+\delta R_b^H=
R_b^{RSM}\,\left({\delta\Gamma_b^{MSSM}\over \Gamma_b^{RSM}}
-{\delta\Gamma_h^{MSSM}\over \Gamma_h^{RSM}}\right)\,.
\label{eq:dRbMSSM}
\end{equation}
 Therefore, the question arises on whether the extra
contributions from the MSSM with respect to the RSM--the MSSM being at present
the most predictive framework for physics beyond the SM-- can solve or at least
soften this conflict between theory and experiment.
We feel that this issue
is important enough for the present and future
credibility of the MSSM to deserve detailed
studies from  different points of view.
In particular we reconsider it within the
context of our fully fledged computation of electroweak SUSY one-loop
corrections to $\Gamma_Z$ presented in Part\, I. In our
approach we extend former calculations\,\cite{b1005,b1GHVR,b1BF5} by
including the full MSSM corrections, not only to the partial width
$\Gamma (Z\rightarrow b\bar{b})$ but also to all quark channels contributing to
$\Gamma (Z\rightarrow {\rm hadrons})$. We treat the Higgs sector of the
MSSM at the one-loop level. Furthermore, our calculation is not just
a leading order calculation projecting specific contributions from
$m_t$-dependent and/or large $\tan\beta$ Yukawa couplings\, ,
but an exact one-loop
calculation including both gauge and Yukawa
couplings on equal footing and
for arbitrary values of $\tan\beta$.  This will be necessary to find out a
reduced
interval of allowed values for $\tan\beta$ where to cure or at least
to alleviate the above discrepancy.
For completeness, we also include
for each $q\bar{q}$ channel the
contribution from the terms
$\nabla_{U,Q}^{SUSY}$ on  eq.(23) of Part\, I, which
in particular involve the full $\Delta r^{MSSM}$. These contributions do not
completely cancel in the ratio $R_b$.

Although the ratio (\ref{eq:Rb}) is practically independent of the Higgs mass
in the SM,
it turns out that the {\sl additional} Higgs contributions
in the MSSM could play an important role, due to enhanced Yukawa couplings.
 To this aim, as already advertised, a first step is called for; namely,
the computation of the quantity $\delta\Gamma_Z^H$ on eq.(\ref{eq:Tshift}).
 Some comments on previous work in this direction are in order.
 The Higgs vertex corrections for the $b\bar{b}$-channel
were first computed in Refs.\,\cite{b1010} and \cite{b1011}. In the former,
extreme values of the Yukawa couplings were used and the small oblique
contributions were neglected; in the latter, the non-oblique corrections were
considered in detail in the
general unconstrained two-Higgs-doublet-model (2HDM) for the $b\bar{b}$ and
$\tau^+\tau^-$ modes and
the universal part was dealt with using a large mass splitting approximation.
(There are some disagreement in the numerical results between these
two references.).
We have nonetheless redone ourselves the entire calculation without any
of the aforementioned
approximations, neither in the treatment of the universal nor in that of the
non-universal parts. We perfectly agree with the numerical results
of ref.\cite{b1011} for the general 2HDM,
but, as noted, use is made of the (one-loop) mass relations
in the Higgs sector of the MSSM\,\cite{b1009}.
In this way we may
assess the relative importance of
$\delta\Gamma_Z^H$ as a part of
the total radiative shift (\ref{eq:Tshift}).
The leading one-loop effects on the Higgs
sector can be extracted from the general formulae of
Ref.\cite{b1009} and one finds the following mass spectrum:
\begin{eqnarray}
m_{H^{\pm}}^2 &=& m_{A^0}^2+M_W^2- {1\over 4}\,\omega_t\,{M_W^2\over
m_t^2}\,,\nonumber\\
m_{H^0,h^0}^2 &=& {1\over 2} \left(\right. m_{A^0}^2+M_Z^2+\omega_t \pm
\left[\right. (m_{A^0}^2+M_Z^2)^2+\omega_t^2\nonumber\\
&-& 4\,m_{A^0}^2\,M_Z^2\,\cos^2{2\beta}
+2\,\omega_t\,\cos{2\beta}\,(m_{A^0}^2-M_Z^2)
\left.\right]^{1/2}\left.\right)\,,
\label{eq:MH1L}
\end{eqnarray}
where $m_{A^0}$ is the pseudoscalar mass\,\cite{b1004} and
\begin{equation}
\omega_t={2\,N_C\,\alpha\,m_t^4\over 4\,\pi\,s^2\,M_W^2\,\sin^2{\beta}}
\log\left({M_{SUSY}^2\over m_t^2}\right)\,,
\label{eq:MH1L2}
\end{equation}
with $M_{SUSY}^2=m_{\tilde{t}_L}\,m_{\tilde{t}_R}$.
For $\omega_t=0$,  the tree-level relations of the MSSM Higgs sector are
recovered.
To subtract the RSM effects one just notes that
in the limit $m_{A^0}\rightarrow\infty$, $h^0$ behaves
like the SM Higgs\,\cite{b1004}.

The indispensable formulae for the radiative corrections in the on-shell scheme
are
given in Part\, I and the computational details are displayed in
Ref.\cite{b1GJS},
so we jump right away to the final numerical results. For the present analysis
we
present all our results in the framework of Model\, I as defined in Part\, I.
The reason is simply that the SUSY spectrum from
Model\, II has no chance to solve by itself  the
``$R_b$ crisis'', since the corresponding sparticles
are too heavy (see, however, later on).
To start with, we display for completeness\,\cite{b1204} in Figs.1a-1b
the quantity $\delta M_W^H$, i.e. the {\sl additional}
Higgs corrections to $M_W$ with respect
to the RSM both for the tree-level and for the one-loop Higgs sector,
where in the latter case we have taken $M_{SUSY}=1\,TeV$.
This allows direct comparison of the Higgs effects
with the genuine SUSY corrections $\delta M_W^{SUSY}$, whose
study we have presented in Ref.\cite{b1012}
 \footnote{See also the parallel study of ref.\cite{b1CHAN}.} .
 In particular, note that there is a large negative correction (Fig.1b)
for small values of $\tan{\beta}$ and of $m_{A^0}$ as compared to the
tree-level correction (Fig.1a). For sfermions and
charginos of ${\cal O}(100)\,GeV$,
this correction could compensate in part the positive genuine
SUSY effect from the sparticle spectrum of Model\, II,
though it
represents only a small fraction of the total SUSY correction in Model\, I
(cf. Figs.1,2 of Ref.\cite{b1012})
As a matter of fact, the one-loop Higgs sector gives,
unlike the tree-level case,
a correction to $M_W$ which is mostly negative and
non-negligible for $m_{A^0}<100\,GeV$. Thus
the one-loop relations (\ref{eq:MH1L}) may help to distinguish between the
radiative
corrections from the Higgs sectors of the MSSM and of the SM.

The extra effects from the one-loop relations (\ref{eq:MH1L}), although
potentially important for $M_W$, have a limited influence on
the corrections to the partial widths of the $Z$ into fermions. They have
essentially negligible repercussion on the propagator corrections, which were
already very small. Notwithstanding,
for the $b\bar{b}$ channel, they may in some cases noticeably shift
the non-oblique corrections, which are overwhelming with respect to the
oblique contributions.
In general the Higgs effects
can be important only for those channels where enhanced Yukawa couplings may be
involved (cf. eqs.(32),(33) of Part\, I).
Thus we plot  on Figs.2a-2b the full quantity (\ref{eq:Tshift}) for the
$b\bar{b}$, $\tau^+\tau^-$ and $\nu_{\tau}\bar{\nu}_{\tau}$ channels.
The plots include
all sorts of oblique and non-oblique effects from sparticles and Higgses.
 In particular, the  SUSY vertex contributions to the
$b\bar{b}$ channel were already
considered in Refs.\cite{b1GHVR} and \cite{b1BF5} in the Yukawa
coupling approximation.
We have checked that in this limit we are in good
agreement with the numerical plots provided by
the latter reference both for charginos and for neutralinos.
For non-extreme values of $\tan\beta$, the gauge parts of the SUSY
contributions
are non-negligible in front of the Yukawa couplings and have to be
included too \footnote{The structure of the
fermion-sfermion-chargino/neutralino coupling,
including both gauge and Yukawa couplings in a general mass-eigenstate basis,
is given e.g. in eqs.(18)-(19) of Ref.\cite{b1GJSH}. Detailed plots in
$(\mu,M)$-space accounting for the full corrections are provided
in Ref.\cite{b1GJS}. }.
 Remarkably enough, the intermediate $\tan\beta$ region
will be essential to the analysis of $R_b$ in
the MSSM for large $m_{A^0}$, as will be shown below. The differences
introduced
by the MSSM Higgses can be appreciated on comparing Figs.2a-2b of this paper
with Figs.4a-4b of Part\, I. Indeed, for pseudoscalar masses in the range
$20\,GeV\stackrel{\scriptstyle <}{{ }_{\sim}}m_{A^0}
\stackrel{\scriptstyle <}{{ }_{\sim}} 60\,GeV$
there is a substantial additional, positive, correction
for high values of $\tan\beta$, especially for the $b\bar{b}$ mode.
 For heavier
Higgs masses and/or lower values of $\tan\beta$, the correction becomes
negative.
Asymptotically in $m_{A^0}$ (and very slowly), the $\delta\Gamma_Z^H$-effect
in the three decay modes goes away,
as we would expect of any  MSSM contribution
entailing a departure with respect to the RSM.
To be precise, in that limit the total vertex Higgs correction in the MSSM
should
boil down to the corresponding RSM contribution, which is negligible.
The existence of the ``positiveness region''
 $20\,GeV\stackrel{\scriptstyle <}{{ }_{\sim}}m_{A^0}
\stackrel{\scriptstyle <}{{ }_{\sim}} 60\,GeV$
was already noticed in
Ref.\cite{b1011} in the context of the general 2HDM. However, while in that
framework the large, positive, contributions correspond to a peculiar choice of
the
free parameters of the model, in the MSSM the wellcome effects
 appear automatically from the constrained
structure of the SUSY Higgs potential. Thus the ``$R_b$ crisis'' can naturally
be
solved in the MSSM within the
 ``positiveness region'', as we shall show explicitly.
In fact, part of this region
($m_{A^0}\stackrel{\scriptstyle >}{{ }_{\sim}} 40\,GeV$)
 has not yet been convincingly
excluded by experiment\,\cite{b1002,b1PR1} and we shall take advantage of this
fact in our analysis.

Aiming at a closer study of the ratio $R_b$ in the MSSM,
we plot in Fig.3 contour lines of
$\delta\Gamma^H (Z\rightarrow b\bar{b})$ in the
$(m_{A^0}, \tan\beta)$-plane.
The extremely slow decoupling of the negative contribution to
 $\delta\Gamma^H (Z\rightarrow b\bar{b})$
for $m_{A^0}\rightarrow\infty$
is also manifest here.
It is worth noticing from Fig.3
that the large, positive, genuine SUSY contribution from
$\delta\Gamma_Z^{SUSY}$ in the $\tan\beta<1$ region (which
we remarked in Part\, I) turns out to be cancelled and even overridden by
the big, negative, contributions from $\delta\Gamma_Z^H$
over a wide range of $m_{A^0}$.
The upshot is that the total MSSM correction (\ref{eq:Tshift})
in the $\tan\beta<1$ region is negative, contrary to naive
expectations from the analysis of $\delta\Gamma_Z^{SUSY}$ alone.
Treating the Higgs sector at 1-loop gives differences in
$\delta\Gamma^H (Z\rightarrow b\bar{b})$ which can be of order of $-1\,MeV$
with respect to the corrections from the Higgs sector at tree-level.
The extra correction basically comes from the vertices, only in the region
around $m_{A^0}=90\,GeV$ and for large $\tan\beta$.
These differences, small as they are,
are of the same order of magnitude--and opposite in sign--to
the typical genuine SUSY effects on the leptonic modes  (cf. Fig 4b of Part\,
I),
and therefore they could result in some cancellation at the level of the total
quantum correction to $\Gamma_Z$. In general we find that
the one-loop effects on the Higgs
masses have little impact on $\Gamma_Z$.

Next we analyze numerically, and in a systematic way,
 the possible solutions to the ``$R_b$ crisis'' both in the
``intermediate Higgs mass range'' ($40\,GeV\stackrel{\scriptstyle <}{{
}_{\sim}}
m_{A^0}\stackrel{\scriptstyle <}{{ }_{\sim}} 70\,GeV$) and
in the ``heavy Higgs mass range''($m_{A^0}> 70\,GeV$).
We start from the latter,
which has already been addressed in
the literature from a different approach\,\cite{b1005}.
Here the MSSM might find itself
in deep water and we have to struggle
a lot more to rescue it from wreckage.
Simple inspection of Fig.3, combined with Figs.2,4 of Part\, I,
suggests that if the ``$R_b$ crisis'' has any chance to be solved in this range
it has to be handled in the ``Higgs desert''
 $2\stackrel{\scriptstyle <}{{ }_{\sim}}\tan{\beta}
\stackrel{\scriptstyle <}{{ }_{\sim}}40$,
where indeed the negative effects from Higgses are practically non-existent,
say less
than $1\,MeV$ (in absolute value), as compared to the typical SUSY corrections
to
the $q\bar{q}$ channels. Above and below a well-defined
band in the $(m_{A^0},\tan\beta)$-plane,
the Higgs corrections are negative definite, sizeable enough, and thus
responsible
for a lower as well as for an  upper limit on $\tan\beta$.
It is one of the main purposes of this work to
show that the effective range of admissible values for $\tan\beta$ can
still be drastically reduced.

When analyzing the extra positive corrections from SUSY to $R_b^{RSM}$ in the
heavy Higgs mass range, we immediately realize
that the main effect comes from the non-oblique
contributions $\delta\Gamma^{SUSY}(Z\rightarrow b\bar{b})$ from
the genuine SUSY sector. This contribution becomes relevant
provided one of the chargino and stop masses is light enough.
In this respect, let us insist on the possibility, not yet ruled out
experimentally
in a compelling way, that the lightest
 R-odd partner of the top quark, $\tilde{t}_1$,  could be much lighter than the
other squarks (even $< M_Z/2$)\,
\cite{b1PR1,b1STE,b1SJA}. Thus since heavy stops are disfavoured in this case,
it follows that the parameter $M_{SUSY}$ on eq.(\ref{eq:MH1L2}) is much smaller
than $1\,TeV$ and hence
the one-loop relations are indistinguishable from the tree-level
ones.

Clearly, an efficient computer code facing a
systematic exploration of positive contributions to $\delta R_b^{SUSY}$
( more specifically, contributions
capable of restoring the quantity (\ref{eq:RbMSSM})
within $1\sigma$ of the experimental result)
is what is needed. The scatter plot method of Ref.\cite{b1005} is one example.
However, in that reference, $\tan\beta$ and $m_{A^0}$ were definitely
fixed at just a couple of values and only the
rest of the parameters were varied.
In our case, we use a simple and straightforward ``lattice'' method in which we
include both of them as additional parameter axes.
This will prove very useful
to explore the range of allowed values for $\tan\beta$ and $m_{A^0}$ .
Thus we first set up  ``seed intervals'' over all SUSY parameter
axes and endow them with a reasonably fine subdivision
in order to generate a sufficiently large number of candidate points
(above $10^8$). Any of such points is defined by a $8$-tuple
\begin{equation}
(\tan\beta, m_{A^0}, M, \mu, m_{\tilde{\nu}}, m_{\tilde{u}}, m_{\tilde{b}},
M_{LR})\,.
\label{eq:tuple}
\end{equation}
Here  $m_{\tilde{\nu}}$ is the sneutrino mass, which enters through the
oblique corrections,  and
$m_{\tilde{u}}$ stands for
the common mass of the $T^3=+ 1/2$ squark components of the two first
generations.
However, as announced, we shall treat
the third squark generation $(\tilde{t},\tilde{b})$ apart.
In particular, we consider the effect of L-R mixing for the $\tilde{t}$ squarks
and parametrize the stop mass matrix in the usual way\,\cite{b1001}
\begin{equation}
{\cal M}_{\tilde{t}}^2 =\left(\begin{array}{cc}
M_{\tilde{b}_L}^2+m_t^2+\cos{2\beta}({1\over 2}-
{2\over 3}\,s^2)\,M_Z^2
 &  m_t\, M_{LR}\\
m_t\, M_{LR} &
M_{\tilde{t}_R}^2+m_t^2+{2\over 3}\,\cos{2\beta}\,s^2\,M_Z^2
\end{array} \right)\,,
\label{eq:stopmatrix}
\end{equation}
where we have used the fact that $SU(2)_L$-gauge invariance requires
$M_{\tilde{t}_L}=M_{\tilde{b}_L}$ and thus the first entry of this matrix can
be written in terms of the sbottom mass
 $m_{\tilde{b}_L}$ (cf.eq.(25) of Part\, I), already included in
(\ref{eq:tuple}).
To illustrate the
effect of the mixing it will suffice to choose the soft SUSY-breaking mass
$M_{\tilde{t}_R}$ in such a way that the two
diagonal entries of  ${\cal M}_{\tilde{t}}^2$ are equal--
the mixing angle is thus fixed at $\pi/4$-- and the remaining free parameter,
$M_{LR}$, is just the last component of the $8$-tuple (\ref{eq:tuple}).
 For the
mixing parameter, however, we have the proviso
\begin{equation}
M_{LR}\leq 3\,m_{\tilde{b}}\,,
\label{eq:MLR}
\end{equation}
which roughly corresponds to a well-known
necessary, though not sufficient, condition to avoid false vacua, i.e. to
guarantee that the $SU(3)_c\times U(1)_{em}$ minimum is the absolute one
\,\cite{b1VACU}.
With all the parameters of the $8$-tuple defined, the ranges that have been
effectively explored for each one of them are the following:
\begin{equation}
\begin{array}{cccc}
0.7<\tan\beta<70 & & 40\,GeV<m_{A^0}<150\,GeV\\
0< M<250\,GeV &  &  -200\,GeV< \mu<200\,GeV \\
50< m_{\tilde{\nu}}<500\,GeV &  & 90< m_{\tilde{u}}<500\,GeV  \\
90< m_{\tilde{b}}<500\,GeV &  &  |M_{LR}|<3 m_{\tilde{b}}\,. \\
\end{array}\
\end{equation}
The final intervals recorded here
are sufficiently stable against progressive
stretching.  As a matter of fact, they
are the result of a number of
consecutive widenings of original, narrower, seed intervals until clear
stabilization was achieved.

Collecting the previous conditions, all SUSY masses are well determined
within the framework of Model\, I.
Obviously, our analysis has unequal sensitivity
to the $8$ parameters in (\ref{eq:tuple}), and this has
been taken into account in the number and distribution of points assigned to
the
various axes. Furthermore,
in order to proceed in an efficient way
(i.e. without wasting a lot of CPU time on obviously sterile points)
we first select (``flag $1$'')
a subset of points (\ref{eq:tuple})
that give rise to at least one light chargino, one light neutralino and
one light stop. For example, a typical setting
would be to require (using the notation of Part\, I)
that there exists at least one index triad  $(i,\alpha, a)$ such that
\begin{equation}
48\,GeV<M_{\Psi^{\pm}_i}<60\,GeV;\ \ M_{\Psi_{\alpha}^0}>20\,GeV;\ \
\ \ 45\,GeV<m_{\tilde{t}_a}<60\,GeV\,.
\label{eq:flag1}
\end{equation}
 Finding flag-$1$-successful points is trivial and very little time consuming.
However, once they are found, the points enter the full computer
flow evaluating the radiative corrections  and
a massive numerical analysis is required to
ascertain, among the many combinations of SUSY parameters that passed
flag $1$ (several millions), those combinations (``flag $2$'')
that fall within the experimental
$1\sigma$ range for $R_b^{MSSM}$ both for
the results (\ref{eq:RbSM2}) and (\ref{eq:RbSM1}) corresponding, respectively,
to plugging or unplugging the CDF limits on $m_t$.
Points that successfully
pass the two flags are to be called ``admissible points''.
Whenever one such point is found, our code projects the corresponding
value of $\tan\beta$ and in this way we are able to
generate a range of admissible values for this
parameter, if any. In particular, using this procedure for $m_t$ within CDF
limits
no point was found for stop masses in the range (\ref{eq:flag1}).
Only for $m_{\tilde{t}}$ well
below $45\,GeV$ a small set of admissible points was detected
at the single value $\tan\beta=4$.
Unavoidably one is forced to go beyond $1\sigma$ to be able to generate
admissible points for $m_{\tilde{t}}>45\,GeV$;
for example, the range $2<\tan\beta<11$ is allowed at $1.25\sigma$.

On the other hand, in the CDF-unrestricted case (\ref{eq:RbSM1}),
we find admissible points already at $1\sigma$ in the range
$2<\tan\beta<10$.
 All of them reach the experimental bounds
for $R_b^{\rm exp}$ ``tangentially'' from below, as shown in Fig.4. In this
figure,
which is rather laborious to compute,
we plot the {\bf maximal} contributions to $R_b^{MSSM}$ as a function of
$\tan\beta$
when all the parameters of the $8$-tuple (\ref{eq:tuple}) are varied with
$m_{A^0}$ in the
heavy Higgs mass range. It should be mentioned that the
alleged  ``tangential solutions'',
which are exclusively associated to light charginos and stops,
are compatible with the experimental bound
on the total width (cf. eq.(2) of Part\, I), as we have checked explicitly.
These solutions are obtained (automatically by our code)
by picking points very close to the
boundary of the allowed region in the $(M,\mu)$-plane for each $\tan\beta$.
 Indeed, near the boundary, the total SUSY contribution
to the $Z$-width is minimum (cf. Fig.1 of Part\, I) while at the same time
$R_b^{MSSM}$ is maximum. The reason for this ambivalent behaviour
is that the large, negative, self-energy corrections from
the ``ino'' sector near that boundary practically cancel
in the difference (\ref{eq:dRbMSSM}) while the positive SUSY vertex corrections
to
$\Gamma(Z\rightarrow b\bar{b})$ are maximum.

Finally, we face systematically the computation of $R_b^{MSSM}$ in the
intermediate Higgs mass range
$40\,GeV\stackrel{\scriptstyle <}{{ }_{\sim}} m_{A^0}
\stackrel{\scriptstyle <}{{ }_{\sim}}70\,GeV$ for $m_t$ within CDF limits.
 Here we wish to show that the
``$R_b$ crisis'' may comfortably be solved in the MSSM
for any SUSY spectrum above the current phenomenological bounds (hence
without resorting to too light stops and charginos), provided $\tan\beta$ is
large
enough. To this end we first plot in
Fig.5a the quantity
$R_b^{RSM}+\delta R_b^{H}$ versus $\tan\beta$ for various pseudoscalar masses
in the aforementioned range. This situation corresponds to $R_b^{MSSM}$ with a
SUSY
spectrum fully decoupled (cf. eq.(\ref{eq:dRbMSSM})).
Since in the range under consideration the Higgs contribution
to the $b\bar{b}$ mode is large, we have to be careful in dealing with  $\delta
R_b^{H}$
by at the same time keeping an eye on the corrections to the total width.
Thus in computing Fig.5a we have imposed the
condition that the total width, given by
\begin{equation}
 \Gamma_Z^{MSSM}=\Gamma_Z^{RSM}+\delta\Gamma_Z^{MSSM}\,,
\label{eq:GZB}
\end{equation}
should not exceed to $1\sigma$ the
experimental value $\Gamma_Z^{\rm exp}$ (cf. eq.(2) of Part\, I)
with all the errors (experimental and theoretical)
added in quadrature.
 As a result of this bound, all the curves in Fig.5a are
cut off at some point (some of them beyond the range explicitly shown)
before exiting the allowed
experimental band for $R_b^{\rm exp}$ at $1\sigma$. In spite of the $\Gamma_Z$
bound,
it is clear from Fig.5a that a well defined solution to the ``$R_b$ crisis''
 exists in the shaded area. Therefore, the lower bound
$\tan\beta\stackrel{\scriptstyle >}{{ }_{\sim}} 36$ ensues.

Next we repeat a similar analysis when switching on the sparticle spectrum.
Here the computation is more difficult since we have to perform
a systematic exploration of the parameter space (\ref{eq:tuple}), e.g. using
the lattice method described above. Another complication is that
we have to separate the case of
``light charginos''( defined as those within the mass interval
assumed on eq.(\ref{eq:flag1})) from the case of
``heavy charginos'' ($>60\,GeV$). The reason
for the separate treatment is that in the light
chargino case, as already noticed
when discussing the ``tangential solutions'',
the global SUSY contribution to the $Z$-width is minimum and so
the analysis of $R_b^{MSSM}$ is not interfered by the
bound on $\Gamma_Z^{\rm exp}$. On the other hand, in the heavy chargino case
the
large negative self-energy corrections disappear, hence the SUSY contribution
to
$\Gamma_Z^{MSSM}$ is boosted up significantly
(cf. Fig.1 of Part\, I) and thereupon the analysis
of $R_b^{MSSM}$ becomes severely restricted by $\Gamma_Z^{\rm exp}$.
 In this letter we shall limit ourselves to display
the results corresponding to a heavy chargino case: specifically in
the intriguing situation where they cannot be pair produced at LEP 200, i.e.
$ M_{\Psi^{\pm}_i}\simeq 100\,GeV$.
 Although we shall briefly comment on
the analysis of $R_b$ for light charginos, and also
for ``intermediately heavy'' charginos (viz.\,
$60\,GeV< M_{\Psi^{\pm}_i}<100\,GeV$), we shall defer a detailed exposition of
the
these results for Ref.\cite{b1GJS}.
 The resulting curves for heavy charginos are shown in Fig.5b.
 In this figure, whose numerical computation is highly CPU-time-demanding,
we project the {\bf maximum} contribution to $R_b^{MSSM}$
as a function of $\tan\beta$ when varying all the parameters of the $8$-tuple
(\ref{eq:tuple}) (except $m_{A^0}$, which is fixed for each curve) with the
condition
that the full sparticle spectrum generated lies just out of
the possibilities of pair production at LEP 200. In practice this means that we
required for sfermions and charginos
\begin{equation}
m_{\tilde{l}^{\pm}_a}\,, m_{\tilde{q}_a}
\,,  M_{\Psi^{\pm}_i}\stackrel{\scriptstyle >}{{ }_{\sim}} 100 \,GeV\,.
\label{eq:boundB}
\end{equation}
Again the restriction from $\Gamma_Z^{\rm exp}$ was imposed on the
corresponding theoretical results
(\ref{eq:GZB}). Two novel features emerge as compared to Fig.5a, namely:

a) A solution to the ``$R_b$ crisis'' exists in the shaded area of Fig.5b, but
this time for
$\tan\beta\stackrel{\scriptstyle >}{{ }_{\sim}} 32$, i.e. starting about $4$
units below
the case with the Higgses alone;

b) The upper cut-off from $\Gamma_Z^{\rm exp}$ on the solution curves of Fig.5b
is so stringent that $m_{A^0}$ and $\tan\beta$ become
strongly correlated; for instance, one
reads from Fig.5b that for $m_{A^0}=50\,GeV$ the only
admissible values for $\tan\beta$ lie in the narrow window
$54\stackrel{\scriptstyle <}{{ }_{\sim}}\tan\beta\stackrel{\scriptstyle <}{{
}_{\sim}}58$.

We point out that for intermediately heavy charginos
the correlation between $m_{A^0}$ and $\tan\beta$ is even
higher than in Fig.5b, due to the severe bound from
$\Gamma_Z^{\rm exp}$ on the large vertex
contributions to the $b\bar{b}$ mode. Quite in contrast, in the light chargino
region,
the correlation disappears and the lower bound on $\tan\beta$ diminishes $12$
units
with respect to the previous case, i.e. $\tan\beta>20$\,\cite{b1GJS},
which is still remarkably high.
In the other extreme, namely
for heavier and heavier sparticle
spectrum, one reaches asymptotically the situation in Fig.5a, where
 it is worth noticing that it corresponds in good approximation
to the one expected for
 Model\, II. In fact, remember that this model is characterized by a
rather heavy SUSY spectrum and it
is closely related to the class of MSSM's with radiatively induced
breaking of the gauge symmetry\,\cite{b1001}. From this point of view
the solution to the ``$R_b$ crisis'' in the intermediate Higgs mass regime
is theoretically preferred to
the ``tangential solution'' obtained for heavy Higgses in Model\, I.
Finally, we mention that we have detected only small differences in the
previous results in the case
where the superpartners of the top quark are very heavy ($\simeq 1\,TeV$),
that is to say, we have verified that
the one-loop relations (\ref{eq:MH1L}) do not alter significantly the shape of
the
large Higgs contributions to the $b\bar{b}$ mode
 in the ``positiveness region''.
Completion of our  numerical search for admissible points (\ref{eq:tuple}) in
all
the cases described above
took several hundred hours of direct CPU time in an IBM(RISC/$6000$) and
an ``$\alpha$''-computer (DEC 3000/300 AXP).

To summarize, we have studied the full set of MSSM corrections
(\ref{eq:Tshift})
to the partial widths of the $Z$ boson into fermions in the context of
phenomenological and supergravity inspired models. In particular,
we have specialized our general framework to find out regions of parameter
space where
the MSSM could help to cure an apparent discrepancy between $R_b^{\rm exp}$ and
$R_b^{SM}$-the alleged ``$R_b$ crisis'' in the SM.
Although further, and more robust, experimental information
is needed before jumping into conclusions, the following considerations may
tentatively be put forward in the meanwhile:

 i) In the heavy Higgs mass range ($m_{A^0}> 70\,GeV$),
we basically agree with the early results of Refs.\cite{b1005,b1GHVR,b1BF5},
in the sense that
both a light stop and a light chargino of ${\cal O}(50)\,GeV$ are needed to
try to rescue the MSSM from the impasse. However, in the light of an
extended multiparametric one-loop analysis of both $R_b$ and $\Gamma_Z$
we enlarge the scope of the conclusions as follows:

 ii) On general grounds
we may state that for small statistical fluctuations around the numbers
(\ref{eq:Rb})-(\ref{eq:RbSM2}), in the heavy Higgs mass range
the experimental result $R_b^{\rm exp}$ can only be approached
``tangentially'' (from below) by the MSSM.
In particular, for $m_t$ within CDF limits,
we find very unlikely that the MSSM could account for $R_b^{\rm exp}$ at
$1\sigma$.

iii) If we, instead, base the previous analysis
on the CDF-unrestricted case, eq.(\ref{eq:RbSM1}),
and the top quark mass happens to be around the central value of
the MSSM fit (specifically $m_t=160\,GeV$),
we find admissible points already at $1\sigma$
in the interval $2<\tan{\beta}<10$,
and only in this interval.

iv) As far as the intermediate Higss mass range
is concerned, our main conclusion is that
the Higgs sector of the MSSM could by itself comfortably
solve the ``$R_b$ crisis'' in the ``positiveness region''
$40\,GeV\stackrel{\scriptstyle <}{{ }_{\sim}}
m_{A^0}<60\,GeV$.
A solution also exists in this region if we superimpose on the
Higgs contribution
any SUSY spectrum above the present phenomenological bounds. However,
if the charginos lie in the intermediate range
$60\,GeV\stackrel{\scriptstyle <}{{ }_{\sim}} M_{\Psi^{\pm}_i}
\stackrel{\scriptstyle <}{{ }_{\sim}}100\,GeV$ and $m_t$ is bound within CDF
limits,
then, the previous ``comfortable solution''
is traded for a ``cut-off solution''( which in a sense is also ``tangential'').
A
characteristic feature of this solution is that
the parameters $m_{A^0}$ and $\tan\beta$ become so correlated that once we are
given
one of them the other gets ``predicted'' within only a small margin.
In general, for $m_{A^0}$ in the
intermediate mass range, the solution space always projects
onto a segment of $\tan\beta$
starting approximately at the suggestive value
$\tan\beta= m_t/m_b\simeq 35$, which is still far below the perturbative limit
$\tan\beta\stackrel{\scriptstyle <}{{ }_{\sim}} 70$.

v) If the pseudoscalar Higgs is
heavy enough, the upper bound derived on $\tan\beta$ in the heavy Higgs mass
region
gives little hope for the recent
$t-b-\tau$ Yukawa coupling $SO(10)$
unification models, which tend to favour very large values for that parameter.
However, if the pseudoscalar Higgs is intermediately heavy, then, these models
are definitely the favourest ones from the point of view of $R_b$.

vi) Of the two frameworks that we have explored for the sparticle
spectrum (Models\, I and II), only the more phenomenological one (Model\, I)
could
solve--and only``tangentially''--the ``$R_b$ crisis'' both
in the heavy and in the intermediate Higgs mass region.
Model\, II, instead, has no chance unless
a Higgs in the intermediate mass region is
invoked, in which case the solution would be comfortable (not ``tangential'').
Thus, surprisingly enough, Model\, II,
 which is theoretically more sounded (in the
sense that it is closely related to SUSY GUT's) could be, in our opinion,
the most natural and appealing scenario in spite of being initially rejected
due to
its rather heavy sfermion spectrum.

 In short, we are tempted to believe that a possible
solution to the ``$R_b$ crisis'' within the
MSSM has to do more with the Higgs sector of the model than with its spectrum
of
genuine SUSY particles
\footnote{A place where one could find the reverse situation, i.e.
potentially large effects from the genuine
supersymmetric part of the MSSM while at the same time
rather handicapped effects from the Higgses,
is in the physics of the top quark decay,
as shown in Ref.\cite{b1GJSH}. The large effects (comparable to QCD) arise for
big $\tan\beta$, even admitting heavy squarks and moderately heavy charginos,
i.e.
a situation compatible with a possible MSSM solution to the
`` $R_b$ crisis'' in the intermediate Higgs mass region.}.
Thus, \underline{\bf if} there is a ``$R_b$ crisis'' at all,
LEP 200 should be able to discover a supersymmetric Higgs, otherwise the
MSSM could be in trouble.
We became aware of a preprint by J.D. Wells, C. Kolda and G. L.Kane
(UM-TH-94-23) and a preprint by J.E. Kim and G.T. Park (SNUTP 94-66)
 where similar questions are addressed from a different point of
view.

\vspace{0.5cm}

{\bf Acknowledgements}: One of us (JS) is thankful to Wolfgang Hollik for
useful
discussions and gratefully acknowledges
the hospitality at the Institut f\"ur Theoretische Physik
der Universit\"at Karlsruhe during a visit. He is also grateful to M. Mart\' \i
nez
for helpful conversations on the experimental status of $R_b$.
This work has  been partially supported by CICYT
under project No. AEN93-0474. The work of DG has also been financed by a grant
of
the Comissionat per a Universitats i Recerca, Generalitat de Catalunya.

\newpage
\vspace{2cm}
\begin{center}
\begin{Large}
{\bf Figure Captions}
\end{Large}
\end{center}
\begin{itemize}
\item{\bf Fig.1} (a)  Additional corrections\,
$\delta M_W^H$ from the tree-level
Higgs sector of the MSSM as a function of the pseudoscalar mass; (b) As in
case (a), but for the one-loop Higgs sector with $M_{SUSY}=1\,TeV$.

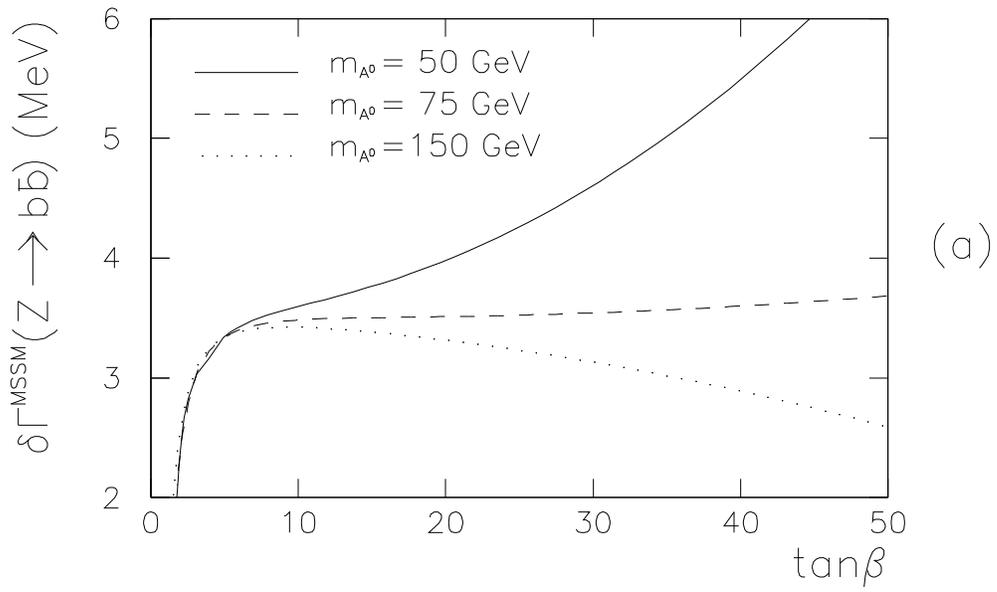
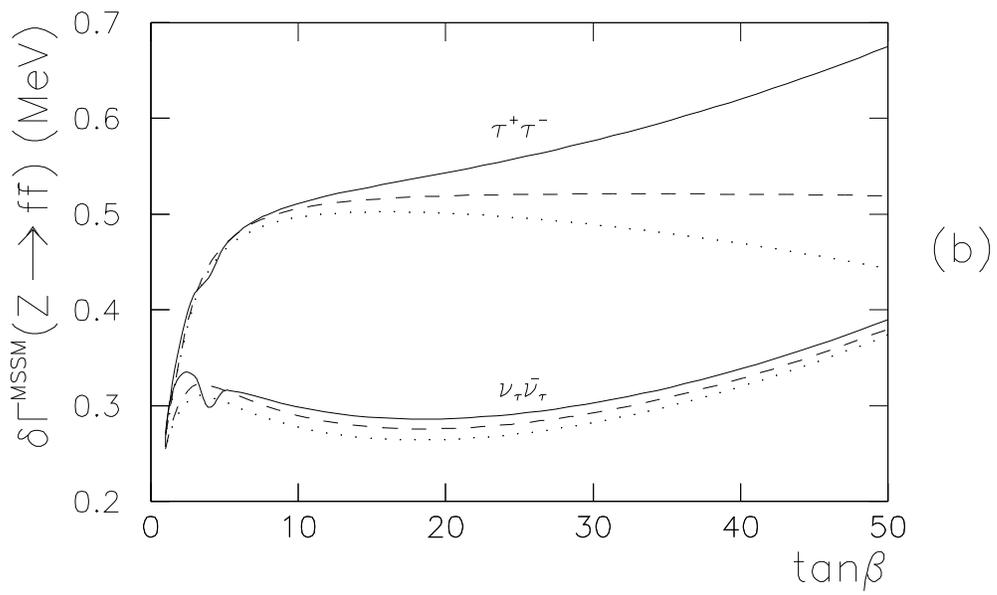
\item{\bf Fig.2} (a) Full correction $\delta\Gamma_Z^{MSSM}$
to the $b\bar{b}$ mode as a
function of $\tan\beta$, for different pseudoscalar masses and the same
spectrum as in Fig.4 of Part\, I; (b) As in case (a), but for the
$\tau^{+}\,\tau^{-}$ and $\nu_{\tau}\bar{\nu}_{\tau}$ modes.

\item{\bf Fig.3} Contour plots of $\delta\Gamma^H(Z\rightarrow b\bar{b})$ in
the $(m_{A^0},\tan{\beta})$-plane.

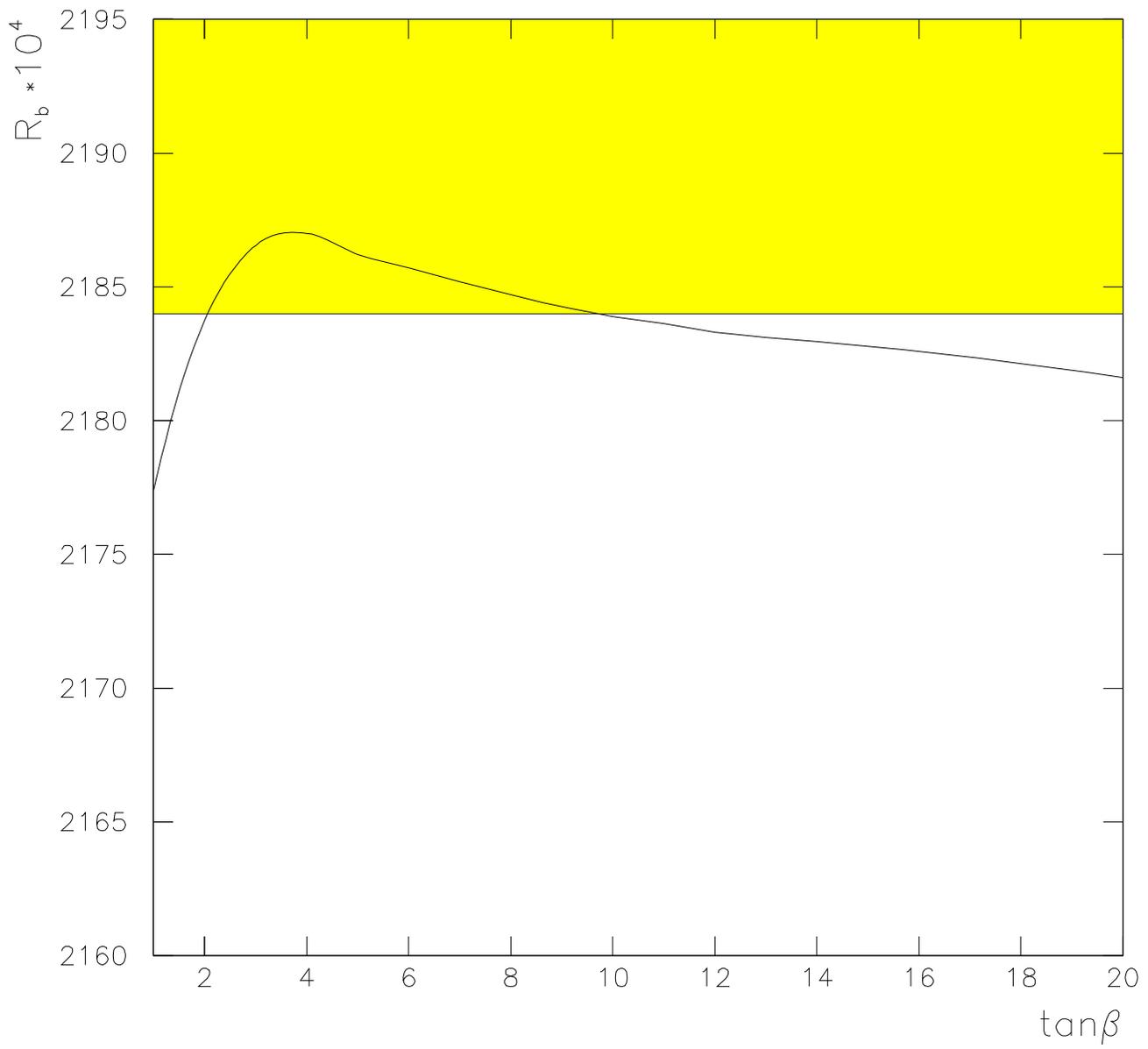
\item{\bf Fig.4} Best ``tangential solution'' in the heavy Higgs mass region
and
for the CDF-unrestricted case.
 The top quark mass is $160\,GeV$ and the sfermion
spectrum is from Model\, I under the optimizing conditions (\ref{eq:flag1}).
The shaded area starts at $R_b^{\rm exp}$ at $1\sigma$.

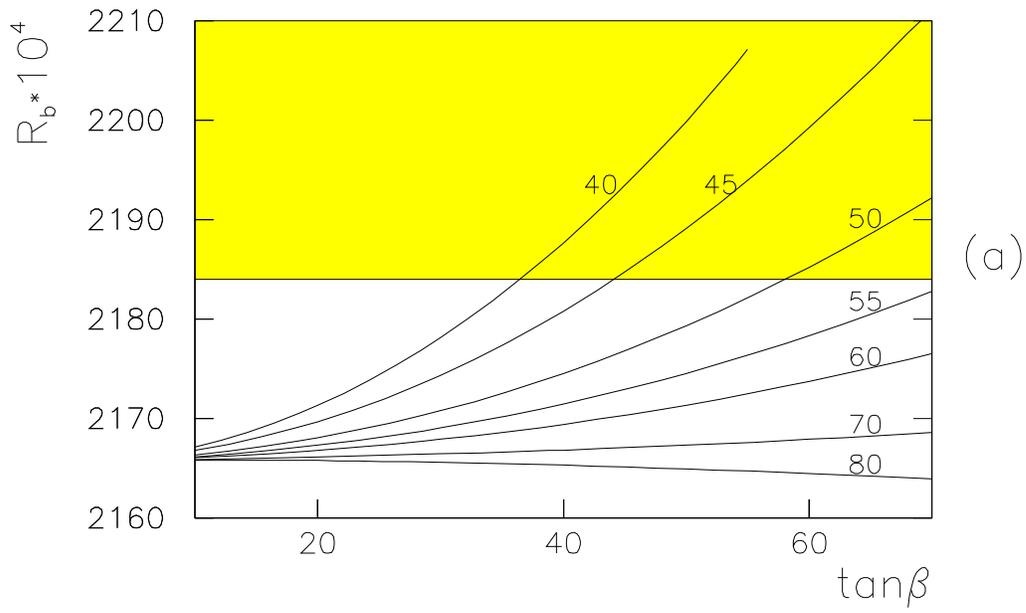
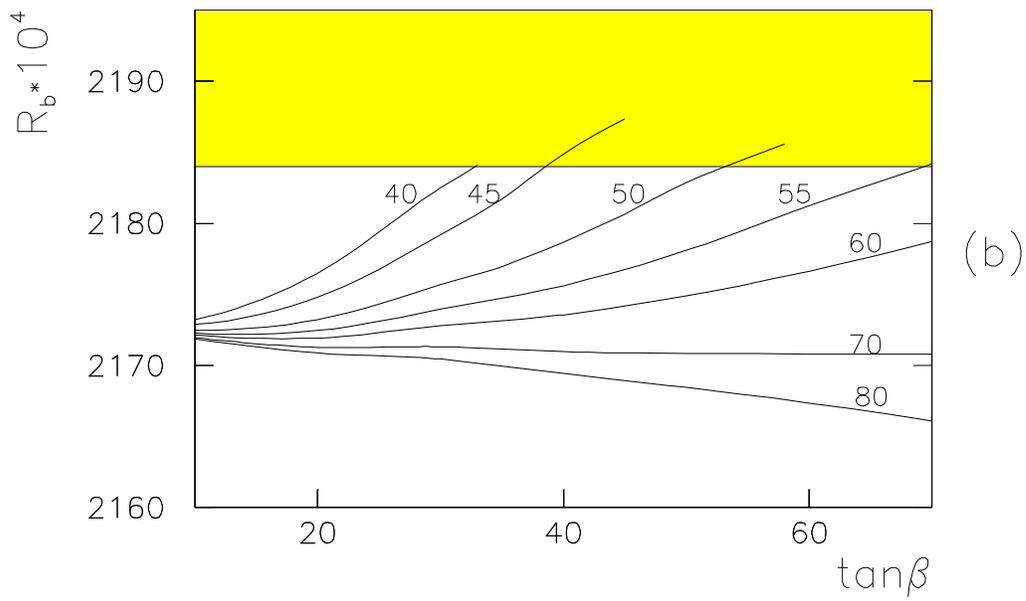
\item{\bf Fig.5} (a) The ``comfortable solution'' for various pseudoscalar
masses (in $GeV$) in the intermediate
Higgs mass region and for a very heavy SUSY spectrum; (b) The ``cut-off
solution'' for the same pseudoscalar masses as before but
for a SUSY spectrum just above the LEP 200 discovery range. In both cases the
shaded area starts at $R_b^{\rm exp}$ at $1\sigma$.

\end{itemize}

\end{document}